\documentclass[11pt,a4paper]{elsarticle}
\usepackage[left=2.54cm,top=2.54cm,right=2.54cm,nohead]{geometry}
\usepackage{setspace}
\usepackage{subfigure}
\newcommand{\picref}[1]{Fig.~\ref{fig:#1}}
\newcommand{\eqnref}[1]{Eq.~(\ref{eq:#1})}

\usepackage{color}
\newcommand{\RA}[1]{#1} % change due to referee #1
\newcommand{\RB}[1]{#1} % change due to referee #2
\newcommand{\RX}[1]{#1} % change for both referees
\newcommand{\RY}[1]{#1} % self-inspired changes
\usepackage{amsmath,amsthm,amsfonts,amssymb}
\usepackage[dvips]{epsfig}
\usepackage[T1]{fontenc}
\usepackage{pslatex}

\title{
Lattice Boltzmann simulations of anisotropic particles at liquid interfaces}
\author{
F. G{\"u}nther$^{*}$, F. Janoschek$^{*}$, S. Frijters$^{*}$, and J. Harting$^{*,**}$\\
Corresponding author: j.harting@tue.nl\\
$^{*}$ Department of Applied Physics, Eindhoven University of Technology, The
Netherlands.\\
$^{**}$ Institute for Computational Physics, University of Stuttgart, Germany.
}
\date{}

\begin{document}
\maketitle

\vskip0.5cm
\centerline{
\begin{minipage}[t]{130mm}
{\bf Abstract:}
Complex colloidal fluids, such as emulsions stabilized by particles with complex
shapes, play an important role in many industrial applications. However,
understanding their physics requires a study at sufficiently large length
scales while still resolving the microscopic structure of a large number of
particles and of the local hydrodynamics.  Due to its high degree of locality,
the lattice Boltzmann method, when combined with a molecular dynamics solver
and parallelized on modern supercomputers, provides a tool that allows such
studies. Still, running simulations on hundreds of thousands of cores is not
trivial. We report on our practical experiences when employing large fractions
of an IBM Blue Gene/P system for our simulations. Then, we extend our model for
spherical particles in multicomponent flows to anisotropic ellipsoidal objects
rendering the shape of e.g. clay particles. The model is applied to a number
of test cases including the adsorption of single particles at fluid interfaces
and the formation and stabilization of Pickering emulsions or bijels.
\end{minipage}
}
\vskip0.5cm

\section{Introduction}
Colloidal particles are highly attractive in the food, cosmetics, and medical
industries to stabilize emulsions or to develop sophisticated ways to deliver
drugs at the right position in the human body. The underlying microscopic
processes of emulsion stabilization with particles can be explained by assuming
an oil-water mixture. Without additives, both liquids phase separate, but
the mixture can be stabilized by adding small particles which diffuse to the
interface and stabilize it due to a reduced interfacial free energy. If for
example individual droplets of one phase are covered by particles, such systems
are referred to as ``Pickering emulsions'', which have been known since the
beginning of the 20th century~\cite{Ramsden1903a,Pickering1907a}. Particularly
interesting properties of such emulsions are the blocking of Ostwald ripening
and the rheological properties due to irreversible particle adsorption at
interfaces or interface bridging due to particle
monolayers~\cite{Binks2006a}.
Recently, interest in particle-stabilized emulsions has led to the discovery of
a new material type, the ``bicontinuous interfacially jammed emulsion gel''
(bijel), which shows an interface between two
continuous fluid phases that is covered by particles. The existence of the
bijel was predicted in 2005 by Stratford et al.~\cite{stratford2005} and
experimentally confirmed by Herzig et al. in
2007~\cite{bib:herzig-white-schofield-poon-clegg}. 

Computer simulations are promising to understand the dynamic properties of
particle-stabilized multiphase flows. However, the shortcomings of traditional
simulation methods quickly become obvious: a suitable simulation algorithm is
not only required to deal with simple fluid dynamics but has to be able to
simulate several fluid species while also considering the motion of the
particles and the fluid-particle interactions. Some recent approaches trying to
solve these problems utilize the lattice Boltzmann method for the description of
the solvents~\cite{succi01}.
The lattice Boltzmann method can be seen as an alternative to conventional
Navier-Stokes solvers and is well-established in the literature. It is
attractive for the current application since a number of multiphase and
multicomponent models exist which are comparably straightforward to implement.
In addition, boundary conditions have been developed to simulate suspended
finite-size particles in flow. These are commonly used to study the behavior of
particle-laden single phase
flows~\cite{ladd01}.  A few groups
combined multiphase lattice Boltzmann solvers with the known algorithms for
suspended particles~\cite{stratford2005,bib:joshi-sun}.  In this paper we
follow an alternative approach based on the multicomponent lattice Boltzmann
model of Shan and Chen~\cite{shan93} which allows the simulation of multiple
fluid components with surface tension. Our model generally allows arbitrary
movements and rotations of rigid particles of arbitrary shape. Further, it
allows an arbitrary choice of the particle wettability -- one of the most
important parameters for the dynamics of multiphase
suspensions~\cite{Binks2006a}. For a detailed introduction to the method see
Ref.~\cite{Jansen2011}, where
our model has been applied to spherical particles
at fluid interfaces. We have presented a thorough validation of the method for
single particle situations and have shown that a transition from a bijel to a
Pickering emulsion can be found by varying the particle concentration, the
particle's contact angle, or the volume ratio of the solvents. Further, we
investigated the temporal evolution of the droplet/domain growth in emerging
Pickering emulsions and bijels.

Modelling colloidal particles as perfect spheres is a strong simplification of
systems appearing in nature. There, the particles are generally not spherical,
but might show geometrical distortions or fully anisotropic shapes, as is, for
example, common for clay particles. As a first step to investigate the
impact of particle anisotropy on the adsorption and stabilization
properties, this paper focuses on ellipsoidal particles.  In addition to
the properties of spheres adsorbed at an interface, in the case of
anisotropic ellipsoidal particles the orientation becomes important and
the process of adsorption is in this case more
complex~\cite{deGraaf2010a}.  Furthermore the anisotropy of the ellipsoids
leads in general to a deformation of the interface. However, an adsorbed
sphere or ellipsoid with a contact angle $\theta=90^\circ$ does not deform
the interface in absence of an external potential such as gravitation.
For multiple particles capillary interactions, which depend on the
distance and the orientation of the particles, become
relevant~\cite{Lehle2008a} and orientational discontinuous phase
transitions of the particles can be found~\cite{Bresme2007a}.
Experimentally it was shown that the number of ellipsoidal particles
required to stabilize a fluid-fluid interface decreases with increasing
particle aspect ratio and that a tip-to-tip arrangement is
dominant~\cite{Madivala2009a}.

The remaining sections are \RY{organized} as follows: In section
\ref{secsimmeth} the simulation method (lattice Boltzmann combined
with molecular dynamics) is illustrated. Since studying
particle-stabilized emulsions demands an exceptional amount of
computing resources we focus on specific implementation details of our
simulation code in section~\ref{secimplementaion}. In \RY{particular}, we
highlight specifically code improvements that allow to harness the
power of massively parallel supercomputers, such as the Blue Gene/P
system JUGENE at J\"ulich Supercomputing Centre with its ability to
run up to $294\,912$ MPI (Message Passing Interface) tasks in
parallel.  The following section reports on simulations of single
particle adsorption of ellipsoidal particles and the formation of
bijels and Pickering emulsions. Finally, we conclude in
section~\ref{secconclusion}.

\section{Simulation method}
\label{secsimmeth}
The lattice Boltzmann method is a very successful tool for modelling fluids in
science and engineering. Compared to traditional Navier-Stokes solvers, the
method allows an easy implementation of complex boundary conditions and---due
to the high degree of locality of the algorithm---is well suited for the
implementation on parallel supercomputers. 
For a thorough introduction to the lattice Boltzmann method we refer to
Ref.~\cite{succi01}. The method is based on a discretized version of
the Boltzmann equation
\begin{equation}\label{eq:LBG}
f_i^c(\mathbf{x}+\mathbf{c}_i,t+1)=f_i^c(\mathbf{x},t)+\Omega_i^c(\mathbf{x},t)
\mbox{,}
\end{equation}
where $f_i^c(\mathbf{x},t)$ is the single-particle distribution function for
fluid
component $c$ after discretization in space $\mathbf{x}$ and time $t$ with a
discrete set of lattice velocities $\mathbf{c}_i$ and 
\begin{equation}\label{eq:BGK_colission_operator}
\Omega_i^c(\mathbf{x},t)=-\frac{f_i^c(\mathbf{x},t)-f_i^\mathrm{eq}(\rho^c(\mathbf{x},t),\mathbf{u}^c(\mathbf{x},t))}{\tau}
\end{equation}
is the Bhatnagar-Gross-Krook (BGK) collision operator.
$f_i^\mathrm{eq}(\rho^c,\mathbf{u}^c)$ is the equilibrium distribution function and $\tau$ is the relaxation time.
We use a three-dimensional lattice and a D3Q19 implementation
($i=1,\ldots,19$).
From \eqnref{LBG}, the Navier-Stokes equations can be
recovered with density $\rho^c(\mathbf{x},t)=\sum_if_i^c(\mathbf{x},t)$
and velocity $\mathbf{u}^c=\sum_if_i^c\mathbf{c}_i/\rho^c$ in the low-compressibility
and low Mach number limit. If further fluid species $c'$ with a single-particle distribution
function $f^{c'}_i(\mathbf{x},t)$ are to be modeled, the inter-species
interaction force
\begin{equation}
  \label{eq:sc}
  \mathbf{F}^c(\mathbf{x},t)=
  -\Psi^c(\mathbf{x},t)
  \sum_{c'}g_{cc'}
  \sum_{\mathbf{x}'}
  \Psi^{c'}(\mathbf{x}',t)
  (\mathbf{x}'-\mathbf{x})
  \mbox{ ,}
\end{equation}
with a monotonous weight function $\Psi^c(\mathbf{x},t)$ for the
effective mass is calculated locally according to the approach by Shan
and Chen and incorporated into the collision term $\Omega_i^c$ in
\eqnref{LBG}~\cite{shan93}. In our case, the coupling strength
$g_{cc'}$ is negative in order to obtain de-mixing and the sum over
$\mathbf{x}'$ runs over all sites separated from $\mathbf{x}$ by one
of the discrete $\mathbf{c}_i$.  Colloidal particles are discretized
on the lattice and coupled to both fluid species by means of a
\RA{moving} bounce-back boundary condition~\cite{aidun98,ladd01}\RA{:
  if $\mathbf{x}$ is part of the surface of a colloid then}
\eqnref{LBG} \RA{for adjacent fluid sites $\mathbf{x}+\mathbf{c}_i$}
is replaced with
\begin{equation}\label{eq:mbb}
  f_i^c(\mathbf{x}+\mathbf{c}_i,t+1)
  =
  f^c_{\bar{i}}(\mathbf{x}+\mathbf{c}_i,t)
  +\Omega_{\bar{i}}^c(\mathbf{x}+\mathbf{c}_i,t)
  +C
  \mbox{ ,}
\end{equation}
where $C=\frac{2\alpha_{c_i}}{c_\mathrm{s}^2}\rho^c(\mathbf{x}+\mathbf{c}_i,t)\mathbf{u}_{\rm
  surf}\cdot\mathbf{c}_i$ is a linear function of the
local particle surface velocity $\mathbf{u}_\mathrm{surf}$ and the direction $\bar{i}$ is
defined via $\mathbf{c}_i=-\mathbf{c}_{\bar{i}}$. 
$\alpha_{c_i}$ and $c_\mathrm{s}$ are constants of the D3Q19 lattice. The particle
configuration is evolved in time solving
Newton's equation in the spirit of classical molecular dynamics
simulations.
As the total momentum should be conserved, an additional force
\begin{equation}\label{eq:mbb2}
%\mathbf{F}(t)=\rho_{\rm init}^c\mathbf{u}_{\rm surf}(\mathbf{x},t)
%\mathbf{F}(t)=\left(2\rho^c(\mathbf{x},t)-
%\frac{1}{6}\rho^c(\mathbf{x},t)\mathbf{u}_{\rm surf}\cdot\mathbf{c}_{i'}\right)\mathbf{c}_{i'}
\mathbf{F}(t)=\left(2f_{\bar{i}}^c(\mathbf{x}+\mathbf{c}_i,t)+C\right)\mathbf{c}_{\bar{i}}
\end{equation}
acting on the
particle is needed to compensate for the momentum change of the fluid caused by
\eqnref{mbb}.
The potential between the particles is a Hertz potential which approximates a
hard core potential and has the following form for two spheres with the same
radius $R$~\cite{bib:hertz}:
\begin{equation}
\phi_H=K_H(2R-r_{ij})^{\frac{5}{2}}\quad\mbox{for}\quad r_{ij}\le2R
\mbox{.}
\end{equation}
$r_{ij}$ is the distance between the two sphere
centers and $K_H$ the force constant. For the simulations which are discussed
later in this text a value of $K_H=100$ is used. In the next step the
potential is generalized to the case of ellipsoids with the parallel radius
$R_p$ and the orthogonal radius $R_o$ by
%following the  method of Berne and Pechukas~\cite{berne72}.
\RB{following the  method which was applied by Berne and Pechukas for the case
  of an intermolecular potential}~\cite{berne72}.
We define
$\sigma=2R$ and $\epsilon=K_H\sigma^{\frac{5}{2}}$
and extend $\sigma$ and $\epsilon$ to the anisotropic case so that
\begin{equation}
\epsilon(\mathbf{\hat{o}}_i,\mathbf{\hat{o}}_j)=\frac{\overline{\epsilon}}{\sqrt{1-\chi^2(\mathbf{\hat{o}}_i\mathbf{\hat{o}}_j)^2}}
\quad\mbox{and}\quad
\label{eq:sigmaaniso}
\sigma(\mathbf{\hat{o}}_i,\mathbf{\hat{o}}_j,\mathbf{\hat{r}}_{ij})
=\frac{\overline{\sigma}}{\sqrt{1-\frac{\chi}{2}(\frac{(\mathbf{\hat{r}}_{ij}\mathbf{\hat{o}}_i+\mathbf{\hat{r}}_{ij}\mathbf{\hat{o}}_j)^2}{1+\chi\mathbf{\hat{o}}_i\mathbf{\hat{o}}_j}+\frac{(\mathbf{\hat{r}}_{ij}\mathbf{\hat{o}}_i-\mathbf{\hat{r}}_{ij}\mathbf{\hat{o}}_j)^2}{1-\chi\mathbf{\hat{o}}_i\mathbf{\hat{o}}_j})}}
\mbox{,}
\end{equation}
with $\overline{\sigma}=2R_o$,
$\chi=\frac{R_p^2-R_o^2}{R_p^2+R_o^2}$, and $\mathbf{\hat{o}}_i$ the
orientation vector of particle $i$.
The scaled potential can be written as
\begin{equation}
\phi_H(\mathbf{\hat{o}}_i,\mathbf{\hat{o}}_j,\mathbf{r}_{ij})
=\epsilon(\mathbf{\hat{o}}_i,\mathbf{\hat{o}}_j)\tilde{\phi}_H\left(\frac{r_{ij}}{\sigma(\mathbf{\hat{o}}_i,\mathbf{\hat{o}}_j,\mathbf{\hat{r}}_{ij})}\right)
\mbox{.}
\end{equation}
$\tilde{\phi}_H$ is a dimensionless function which takes the specific form of
the potential form into account.
In addition to adding the direct interaction described by the Hertz potential we correct
for the limited description of hydrodynamics when two particles come very
close by means of a lubrication correction.  If the number of lattice
points between two particles is sufficient, the lattice Boltzmann
algorithm reproduces the correct lubrication force automatically. The
error that occurs if the flow is not sufficiently resolved can be
corrected by
\begin{equation}
\mathbf{F}_{ij}=\frac{3\pi\mu\RY{R^2}}{2}\mathbf{\hat{r}}_{ij}(\mathbf{\hat{r}}_{ij}(\mathbf{u}_i-\mathbf{u}_j))\left(\frac{1}{r_{ij}-2R}-\frac{1}{\Delta_c}\right)
\end{equation}
in the case of two spheres with radius $R$. We choose a cut-off at $\Delta_c=\frac{2}{3}$ and
$\mathbf{u}_i$ is the velocity of particle $i$.
This equation is generalized to ellipsoids in a similar way as the
Hertz potential using
\eqnref{sigmaaniso}.

\begin{figure}
  \begin{center}
  \includegraphics[width=5.5 cm, height=5.5 cm]{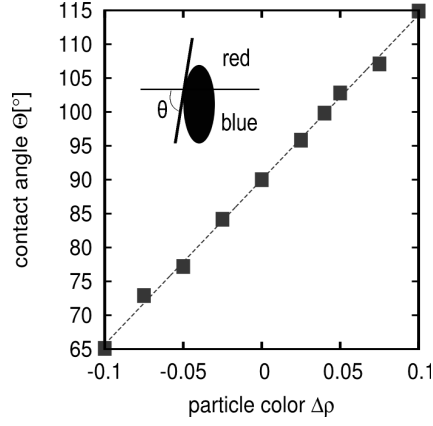}
  \end{center}
  \caption{\label{fig:color_contact-angle_and_contact-angle-definition}
%{\FIXME Axes labels wrong!}
    Relation between the particle color $\Delta\rho$ and the contact angle $\Theta$. A linear relation is
    found: $\Theta=243.2^\circ\Delta\rho+90^\circ$.
    Inset: Definition of the contact angle  $\Theta$ for an ellipsoidal
    particle adsorbed at an interface between two fluids.}
\end{figure}

The force in \eqnref{sc} also includes interactions between lattice
nodes outside of particles with a lattice node inside a particle.
To calculate these interactions the lattice nodes $\mathbf{x}$ in the outer
shell of the
particle are filled with fluid densities
\begin{equation}
\rho_{new}^c=\overline{\rho}^c=\frac{1}{N_{NP}}\sum_{i_{NP}}\rho^c(\mathbf{x}+\mathbf{c}_{i_{NP}},t)
\end{equation}
corresponding to the average over the $N_{NP}$ non-particle lattice nodes
adjacent to $\mathbf{x}$ in directions $i_{NP}$.
We consider a system of two immiscible fluids, which we call red and blue fluid, and in which particles are suspended.
By defining a parameter $\Delta\rho$, the particle color, we are able to
control the interaction between the particle surface and the two fluids
and thus control the contact angle $\Theta$ as it is
defined in the inset of
\picref{color_contact-angle_and_contact-angle-definition}. If $\Delta\rho$
has a positive value, we add it to the red fluid component as
$\rho_{new}^r=\overline{\rho}^r+\Delta\rho$.
Otherwise we add its absolute value to the blue fluid as
$\rho_{new}^b=\overline{\rho}^b+|\Delta\rho|$.
In \picref{color_contact-angle_and_contact-angle-definition} it is
depicted that the dependence of the contact angle on the particle color
can be fitted by the linear relation
\begin{equation}
  \Theta=243.2^\circ\Delta\rho+90^\circ
\mbox{,}
\end{equation}
where the slope depends on the actual simulation parameters. For a more
detailed description of our simulation algorithm the reader is referred
to Ref.~\cite{Jansen2011}.

\begin{figure}
  \centering
  \subfigure[]
  {\label{fig:speedup-nomd}\includegraphics[height=6.3cm]{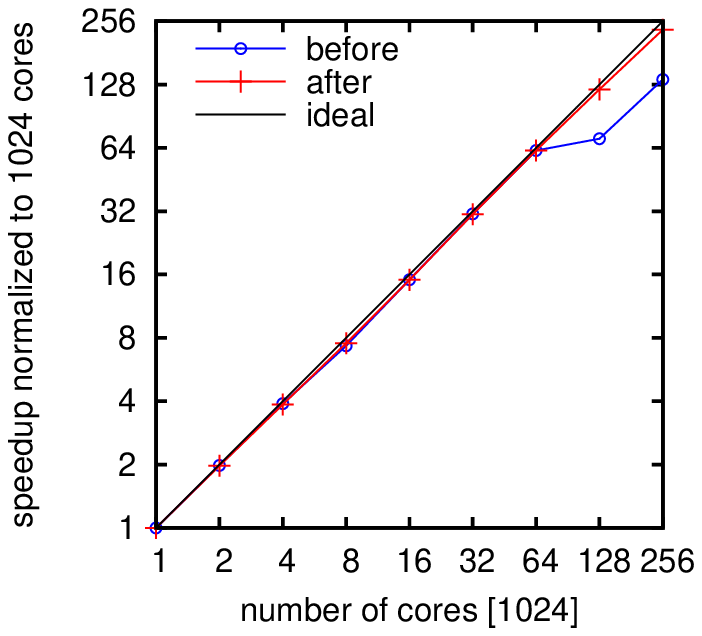}}
\qquad
%  \hfill
  \subfigure[]
  {\label{fig:speedup-md2}\includegraphics[height=6.3cm]{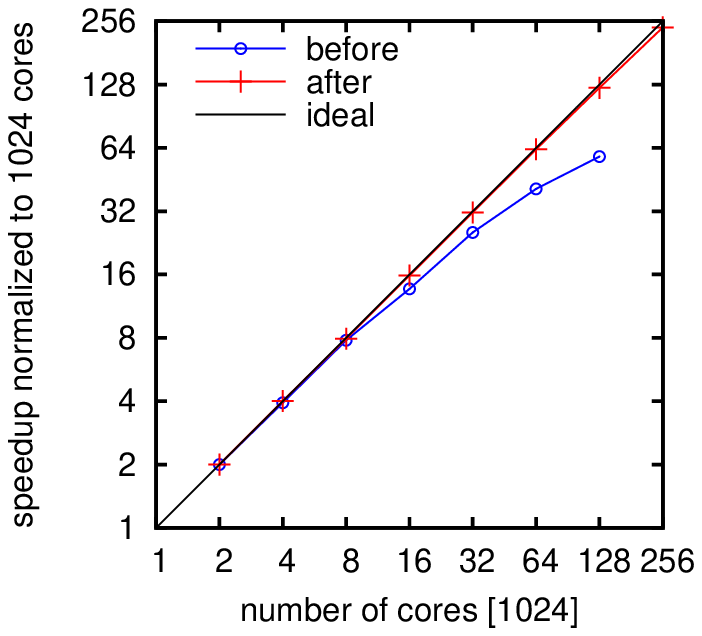}}
  \caption{Strong scaling of LB3D on the Blue Gene/P before and after the
    optimizations. \subref{fig:speedup-nomd} relates to a system with
    only one fluid component so that the effect of matching or mismatching
    topologies of network and domain decomposition can be examined
    better. \subref{fig:speedup-md2} refers to a system with two fluid
    species and suspended particles as they are of interest in this paper.
    The absolute execution times for small
    core counts did not change significantly.}
\end{figure}

\section{Implementation}
\label{secimplementaion}
\RB{Development of our simulation code LB3D started in 1999 as a
  parallel LB solver capable of describing systems of up to three
  fluid species} coupled by Shan and Chen's aforementioned
approach~\cite{shan93,chen00}. In 2008, a---since then severely
extended---parallel molecular dynamics code\RB{~\cite{plimpton95}} was
integrated into LB3D \RB{using a common} 3-dimensional spatial
decomposition scheme\RB{. It is employed here} to implement \RB{the}
colloidal \RB{particles.}

\RX{The} strong locality of the LB equation~\eqnref{LBG} as well as
the short-range interaction forces relevant in colloidal \RX{systems
  generally allow} for an efficient \RX{parallelization. However,
  while several authors presented highly efficient combined
  single-component LB-suspension codes for the massively parallel
  simulation of blood flows on IBM Blue Gene/P
  systems~\cite{clausen10,peters10}, it still is not trivial to
  achieve good scalability on this platform. For example, the network}
employed for \RB{MPI} point-to-point \RB{communication provides}
direct links only between \RB{nearest} neighbors in a
three-dimensional torus. \RB{Allowing} \texttt{MPI\_Cart\_create()} to
reorder process ranks and manually choosing a domain decomposition
\RB{that fits} the known hardware topology \RB{can therefore increase
  the performance significantly at high degrees of
  parallelism. \picref{speedup-nomd} demonstrates this on the basis of
  strong scaling speedup for a system of $1024^2\times2048$
  lattice sites carrying only one fluid species and no
  particles. Consequently, only \eqnref{LBG} demands significant
  computational effort, which makes possible communication bottlenecks
  more visible. On the other hand, JUGENE, the IBM Blue Gene/P system
  at J\"ulich Supercomputing Centre, consisting of $294\,912$ cores,
  brings to light serial parts of a code which would stay
  undetected at lower core counts. \picref{speedup-md2} visualizes the
  effect of parallelizing a loop over the global number of colloidal
  particles present in earlier versions of our particle-fluid coupling
  routines which---in this strong scaling benchmark---had a visible
  effect only when scaling beyond $8192$ cores. The test system of
  $1024^2\times 2048$ lattice nodes contains two fluid species and
  $4\,112\,895$ uniformly distributed particles of spherical shape
  with a radius of 5 lattice units at a volume concentration of
  $20\,\%$ and is thus} very close to the Pickering systems of
interest\RX{. When} \RA{running on $262\,144$ cores, our two-component
  LB-suspension solver achieves an average of $8.97\times10^9$ lattice
  updates per second (LUPS) in total at a decomposition of
  $32\times16^2$ LB sites per core. This corresponds to
  $3.42\times10^4$ LUPS per core compared to $3.66\times10^4$ LUPS per
  core for the minimum core count of $2048$ or a relative parallel
  efficiency of $93.5\,\%$. On average, for $262\,144$ cores,
  $41.8\,\%$ of the computing time is spent on the Shan-Chen force
  evaluation \eqnref{sc}, followed by $29.6\,\%$ for the
  particle-fluid coupling and $18.7\,\%$ for the LB equation
  \eqnref{LBG}. The remaining costs are mainly caused by
  communication}\RX{.}

While the benchmarks above relate to the pure time evolution, actual
production runs require checkpointing and the output of physical
observables to \RB{disk. Using} parallel HDF5 output, we succeed in
storing fluid density fields of $4.6\,\mathrm{GB}$ size for a system
of $1024^2\times1152$ lattice sites and $454\,508$ particles
in $29\,\mathrm{s}$ (on average) when \RB{employing} the whole
system ($294\,912$ cores). This corresponds to the time required to
simulate about $100$ LB steps \RB{and is acceptable since output is
  required less than once per $100$ to $1000$ time steps}.

\RB{Finally, it is important to be aware of possible peculiarities of
  the MPI implementations found on large supercomputing systems. Due
  to their relatively small size in memory we collect and write
  particle configuration data serially on the root process. When
  running on $131\,072$ cores of JUGENE, we encounter a speedup of
  $77$ when employing the specially optimized
  \texttt{MPI\_Allgatherv()} compared to the more intuitive
  \texttt{MPI\_Gatherv()},} at the cost of requiring one receive
buffer per \RB{task}.

\section{Results and discussion}\label{ch:results}
\begin{figure}
  \centering
  \subfigure[]
  {\label{fig:a15h01}\includegraphics[width=0.4\columnwidth]{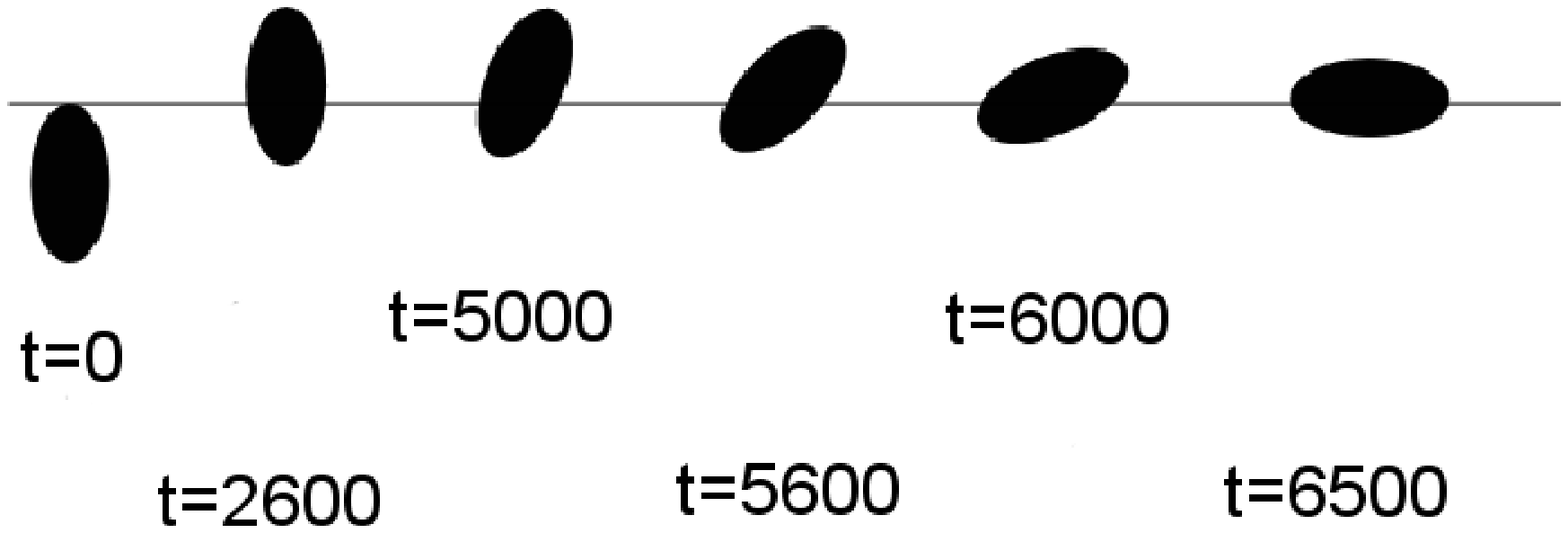}}
  \qquad
  \subfigure[]
  {\label{fig:adsorbmpv}\includegraphics[height=5.5cm]{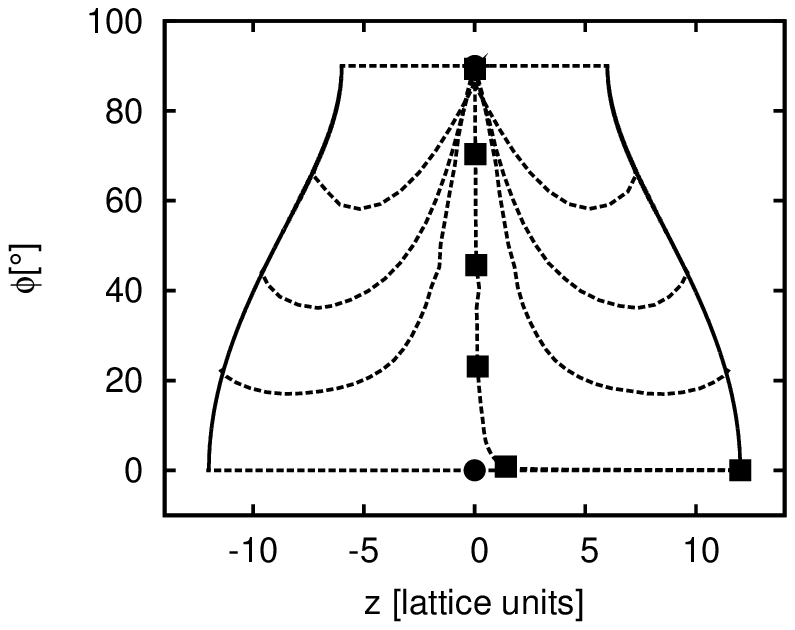}}
  \caption{(a)~Snapshots of the adsorption of an ellipsoid with
aspect ratio $m=2$ and a contact angle $\Theta=90^\circ$ for a starting angle
of $\phi=0.0057^{\circ}$. Every snapshot is related to a square symbol in
\picref{adsorbmpv}. (b)~Adsorption diagram for an ellipsoid with a parallel
diameter $R_p=12$, an orthogonal diameter $R_o=6$, and a contact angle
$\Theta=90^\circ$. $z$ is the distance between the particle center and the
interface, $\phi$ is the angle between the ellipsoid main axis and the
interface normal. The
adsorption trajectories are presented by the dashed lines. The solid lines depict where the particle
would touch an infinitely thin and undeformed interface. The circles
correspond to the stable ($\phi=\frac{\pi}{2}$) and the metastable ($\phi=0$)
equilibrium point. The squares correspond to the snapshots in
\picref{a15h01}.}
\end{figure}
After having discussed the implementation of our program code some of the
results obtained by this program will be presented. We study a system
of two immiscible fluids (called red and blue fluid in the following)
stabilized by prolate ellipsoidal particles. A first step to understand
fundamental physical properties of such systems is the study of the behavior
of a single ellipsoidal particle at a flat fluid-fluid interface.
If the particle touches the interface
the particle is adsorbed \RY{to it}. The binding
energy of the particle is usually of the order $10^{4}k_BT$ so that the adsorption
is an irreversible process.
\RB{This binding energy arises from the fact that an adsorbed particle reduces the
interfacial area between two fluids with repulsive interactions.
The adsorption was studied in Ref.~\cite{deGraaf2010a} using a free energy
method. There, the free energy is calculated for
all particle orientations and particle center-interface distances.
Then, the equilibrium state is found by minimizing the free energy function.
In contrast to our simulations, in the approach given in
Ref.~\cite{deGraaf2010a} the interface is infinitely
thin and undeformable, while we take a finite interface thickness and a interface deformability into account.
}

\RB{
To simulate a single-particle adsorption we use a cubic volume with a side
length of 64 lattice units with two lamellae of different fluids where each
lamella has a thickness of 31 lattice units.
The boundary conditions are periodic in the two directions parallel to the
fluid-fluid interface. Impenetrable walls are installed in the third direction
at $z=1$ and $z=64$ with a thickness of 1 lattice unit each. To keep the interface thin enough the fluid-fluid
interaction parameter between red and blue (\eqnref{sc}) is set to $g_{br}=0.1$.}
Snapshots of the adsorption for a single ellipsoidal particle with a contact
angle $\Theta=90^\circ$ and an aspect ratio $m=\frac{R_p}{R_o}=2$ are shown in
\picref{a15h01} for a starting angle $\phi=0.0057^{\circ}$. $R_p$ and $R_o$
are the ellipsoidal radii parallel and orthogonal to the symetrie axis. It can
be observed that at the beginning of the simulation (first 2600 time steps in
the present example) the particle moves in the direction of the interface
without changing its orientation significantly.  Thereafter ($t=2600...6500$
steps) the orientation changes from $\phi\approx0^{\circ}$ to
$\phi\approx90^{\circ}$ and the particle reaches its equilibrium position.

Each snapshot in \picref{a15h01} is related to a square symbol in
\picref{adsorbmpv}.  Here, $z$ is the distance between the particle center and
the interface, $\phi$ is the angle between the ellipsoid main axis and the
interface normal. The solid lines correspond to the points where the particle
just would touch an infinitely thin and undeformed interface. The dashed lines
correspond to the adsorption trajectories and the two equilibria are
illustrated by \RY{circles}.  The upper \RY{circle} (at $\phi=90^\circ$)
corresponds to the stable point being in relationship with the global free energy minimum
whereas the lower one (at $\phi=0^\circ$) shows a
metastable point.
The value of the $z$ coordinate of the stable and metastable point depends on
the contact angle. In the example shown above both points have the value $z=0$.  If
the adsorption trajectory of the particle starts with $\phi(t=0)=0^\circ$ it
reaches the metastable point. For any other starting angle
($\phi(t=0)\neq0^\circ$) the adsorption trajectory of the particle ends in the
stable point.
The adsorption lines approach attractor lines.
For initial angles much larger than $\phi=0^\circ$ there are two
attractor lines, one on each side. The adsorption trajectories coming from the
respective sides approach this attractor line.  There is a third attractor line
at $z=0$. The adsorption trajectories starting at very small angles approach
this attractor line.
Qualitatively comparable theoretical calculations of the adsorption have been presented in
Ref.~\cite{deGraaf2010a}. Quantitative comparisons are not easily possible since the theoretical approach assumes an infinitely thin and undeformable interface, while our diffuse interface simulations take the interfacial deformability into account.

\begin{figure}
  \centering
  \subfigure[]
  {\label{fig:bijel}\includegraphics[width=6.4 cm, height=6.4 cm]{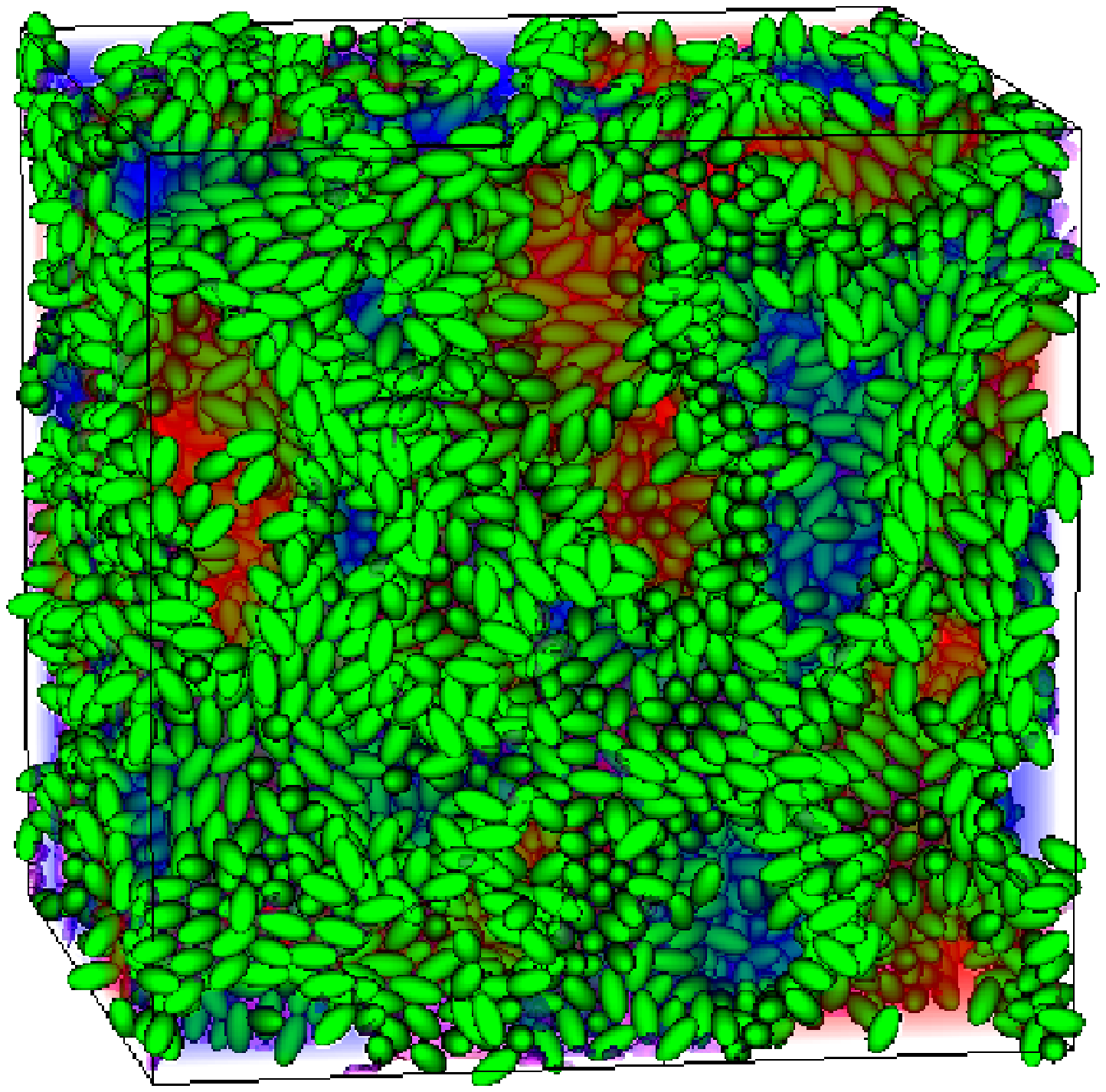}}
  \qquad
  \subfigure[]
  {
  \label{fig:pickemul}\includegraphics[width=6.4 cm, height=6.4 cm]{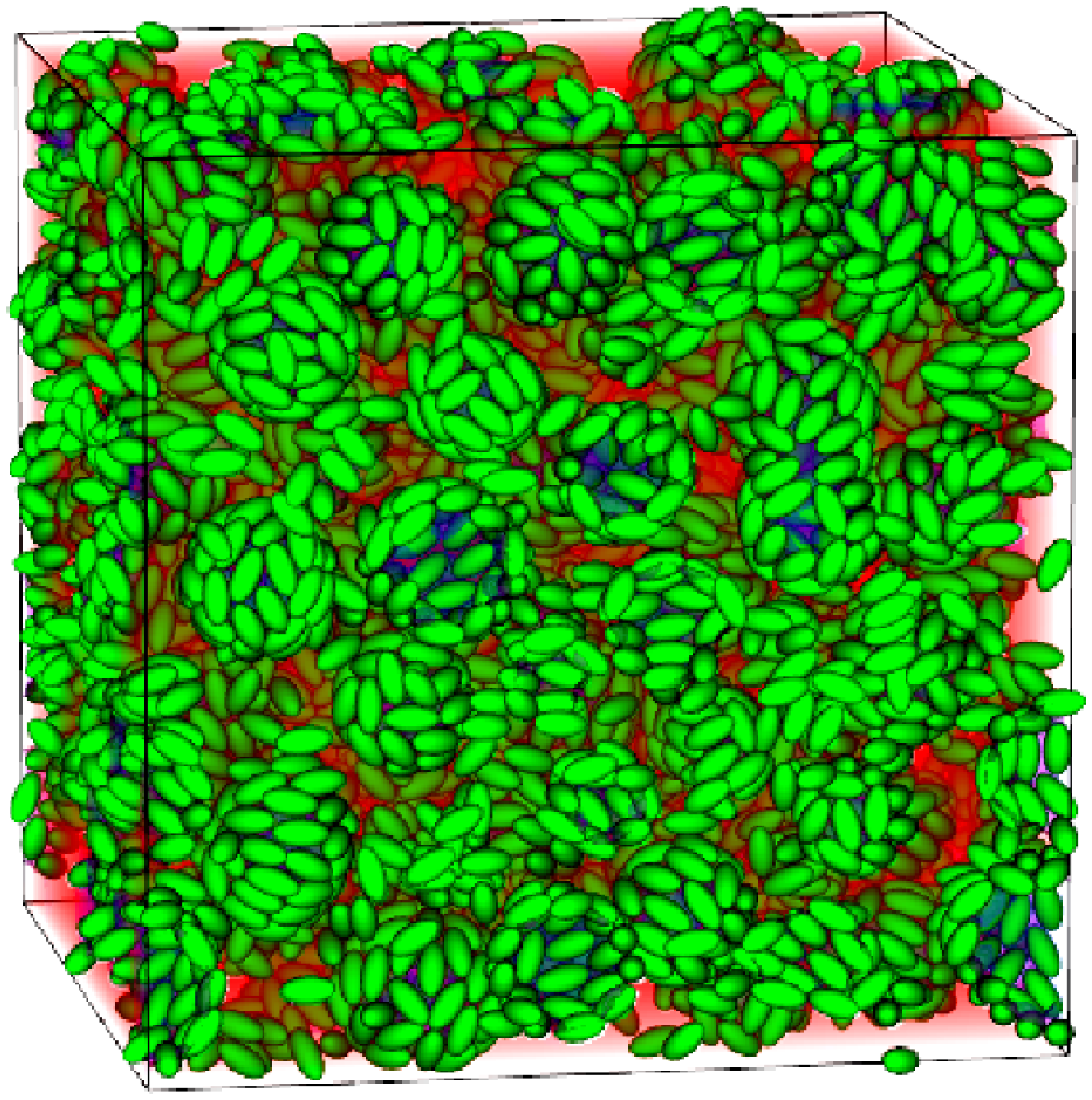}}
  \caption{Two snapshots of emulsions stabilized by particles are
    shown for an aspect ratio $m=2$, a volume concentration of
    $C\approx0.2$ and a contact angle
    $\Theta=90^{\RY{\circ}}$. \RY{\subref{fig:bijel}} Bijel for a
    fluid \RY{ratio} of $1:1$. \RY{\subref{fig:pickemul}} Pickering
    emulsion for a fluid \RY{ratio} of $5:2$.}
\end{figure}

In the following we consider emulsions stabilized by a large number of
anisotropic colloidal particles.
\RB{To simulate a bulk system we use a cubic volume with a side length of 256 lattice
units and periodic boundary conditions. To obtain a phase separation the fluid-fluid interaction parameter between red and blue (see \eqnref{sc})
is set to $g_{br}=0.08$. The initial density distributions of the
two fluids are chosen to be random.}

As stated in the introduction we distinguish
between two different phases for these emulsions, the Pickering emulsion and
the bicontinuous interfacially jammed emulsion gel (bijel). The
bijel~\cite{stratford2005} (see \picref{bijel}) consists of two continuous
phases whereas the Pickering emulsion~\cite{Ramsden1903a,Pickering1907a} (see
\picref{pickemul}) consists of droplets of one fluid immersed in a second fluid
phase.  Bijels and Pickering emulsions stabilized by spherical particles have
been investigated in Ref.~\cite{Jansen2011}. In this paper simulation results
for ellipsoids with an aspect ratio of $m=2$ are presented.
The choice of the control parameters decides into which of the two phases the
system evolves. The left plot in Fig.~\ref{fig:phasesandLC} shows the
transition from a bijel to
a Pickering emulsion for an aspect ratio of $m=2$ and a particle volume
concentration of $C\approx0.2$. The two control parameters used for the study
of the phase transition are the fluid ratio and the contact angle $\Theta$.
The squares show the configurations which lead to a bijel whereas the
circles denote a Pickering emulsion.  If the amount of the two fluids
present in the simulation is equal or not too different (e.g. a ratio of $4:3$)
we find a bijel for all considered contact angles. However, if the fluid ratio
is increased we find a Pickering emulsion. For intermediate fluid ratios the
obtained phase depends on the chosen contact angle. For example for a ratio of
$9:5$ we get a bijel for a contact angle of $90^{\circ}$ and a Pickering
emulsion for all higher values of $\Theta$.

To characterize an emulsion the time dependent lateral domain size
$L(t)=\frac{1}{3}(L_x(t)+L_y(t)+L_z(t))$ is calculated. Its Cartesian components ($i=x,y,z$) are defined as
\begin{equation}
L_i(t)=\frac{2\pi}{\sqrt{\langle k_i^2(t)\rangle}},\qquad
\langle k_i^2(t)\rangle=\frac{\sum_{\mathbf{k}}k_i^2(t)S(\mathbf{k},t)}{\sum_{\mathbf{k}}S(\mathbf{k},t)}.
\end{equation}
$\langle k_i^2(t)\rangle$ is the second-order moment of the three-dimensional
structure
function $S(\mathbf{k},t)=\frac{1}{N}|\phi_\mathbf{k}'(t)|$.
$\phi'=\tilde{\phi}-\langle\tilde{\phi}\rangle$ is the Fourier transform of the
fluctuations of the order parameter $\tilde{\phi}=\rho_r-\rho_b$.
\RB{The right plot in \picref{phasesandLC} shows the dependence of the domain
  size on the volume
concentration of particles $C$ for different concentrations between $C=0.08$
and $C=0.24$ for an emulsion with a fluid ratio of $1:1$ and
$\Theta=90^\circ$.

\begin{figure}
  \begin{center}
  \includegraphics[height=5.6 cm]{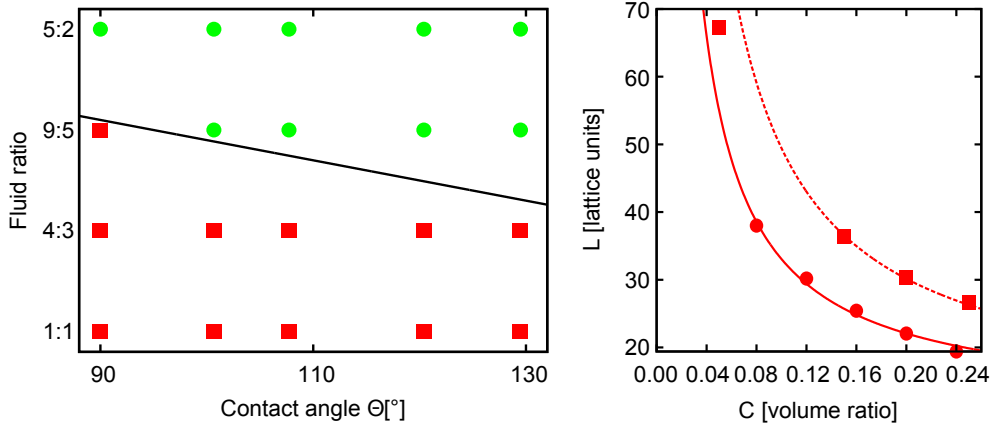}
  \end{center}
  \caption{\label{fig:phasesandLC}
Left plot: Phase diagram demonstrating the transition from a bijel to
a Pickering emulsion. The contact angle $\Theta$ and the ratio between the
two fluids are varied. The squares show the configurations which lead to a
bijel whereas the circles denote a Pickering emulsion. Right plot: Average
domain
size $L$ in equilibrium versus the volume concentration of the particles
\RB{$C$ for a bijel with particles of neutral wettability ($\Theta=90^\circ$).
The plot compares the data for spheres ($m=1$, cubic symbols)~\cite{Jansen2011} with ellipsoids
(here: $m=2$, circles).
The behavior can be
described by $L=\frac{a}{C}+b$ (see solid line ($m=2$) and dashed line
($m=1$)), where $a$ and $b$ are fit parameters.}}
\end{figure}

The results for anisotropic particles (here: $m=2$) and spheres ($m=1$) are compared.
As can be seen in the left inset of \picref{phasesandLC} these parameters for the case of $m=2$
lead to a bijel. For too small volume concentrations such as $C\approx0.04$ there are not
sufficiently many particles available to stabilize the emulsion. The system finally reaches a fully
demixed state.
The simulation data is displayed by circles for $m=2$ and squares for $m=1$ (see \cite{Jansen2011}).
$L$ decreases with an increasing value of $C$. An increasing number of particles
stabilizing the emulsion leads to an increase of the total interfacial area and
thus to smaller domain sizes. The results can be fitted by the
equation $L(C)=\frac{a}{C}+b$~\cite{Jansen2011} (see solid ($m=2$) and
dashed ($m=1$) lines in
the right plot in \picref{phasesandLC}) with values $b\approx11.08$ for $m=2$
and
$b\approx10.85$ for $m=1$. The offset
$b$ is due to the finite system size in our simulation and is
  approximately independent of the particle shape.
% and $a\approx2.20$.
%The factor $a$ depends on the shape of the particles.
%For the case of spheres
%($m=1$), a value of $a\approx3.86$ was found~\cite{Jansen2011} \RB{(see dashed
%line in \picref{ggsfCfit_mr1_3v})}. For
%prolate ellipsoids $a$ decreases with increasing aspect ratio $m$.
The factor $a$ depends on the shape of the particles. For spheres we find
a value of $a\approx3.86$ whereas for $m=2$ we find $a\approx2.20$. 
The right plot in \picref{phasesandLC} shows that the domain size $L$ is
larger for
spherical particles than for the case of anisotropic particles. If a spherical
particle with a radius $R$ and a contact angle $\Theta=90^\circ$ is adsorbed it reduces the
interfacial area by an amount of $A(m=1)=2\pi R^2$. If an anisotropic ellipsoid
with aspect ratio $m$ and the same volume as the sphere is adsorbed and
reaches the stable point (see results of the single-particle adsorption
(\picref{a15h01})) the interfacial area is reduced by an amount of $A(m)=2\pi
R_pR_o=2\pi m^{1/3} R^2$. For the example of $m=2$ we obtain a reduction of the interface area of $A(m=2)\approx2\pi1.26R^2$. Thus, if the particle anisotropy increases, the
occupied area $A(m)$
increases. This means that if the particle number is kept constant and the
anisotropy is increased, the particles can stabilize a larger interface
which leads to smaller domain sizes.
As the structure of a bijel interface is quite
complicated and the interfaces are generally curved it is not possible to
utilize the relations given above to derive an exact dependence of the domain
size $L$ on the particle aspect ratio $m$. However, the qualitative difference of using spheres or ellipsoids as emulsion stabilizers can be well understood.} 

%\picref{dsft_mr1_3v3}
%shows the time dependence of the domain size for different volume
%concentrations $C=0.08$ (highest line), $C=0.12$, $C=0.16$, $C=0.20$ and
%$C=0.24$ (lowest line). Every line corresponds to a circle in
%\picref{ggsfCfit_mr1_3v}.
%The required time for reaching the equilibrium state depends on $C$. For lower
%concentrations ($C=0.08$) it takes about 40000 timesteps whereas for the
%highest presented concentration $C=0.24$ only of the order of 10000
%time steps are required to obtain the final simulation state.
%\begin{figure}
%  \centering
%  \subfigure[]
%  {\label{fig:ggsfCfit_mr1_3v}\includegraphics[width=5.5 cm, height=5.5 cm]{ggsfCfit_mr1_3pbw.ps}}
%  \qquad
%  \subfigure[]
%  {\label{fig:dsft_mr1_3v3}\includegraphics[width=5.5 cm, height=5.5 cm]{dsft_mr1_3pbw.ps}}
%  \caption{(a)~Average domain size $L$ in equilibrium versus the volume
%concentration of the particles $C$ for a bijel with ellipsoids of $m=2$ and
%$\Theta=90^\circ$. The simulation data is displayed by circular points. The
%behavior can be described by $L=\frac{a}{C}+b$ (see solid line), where $a$ and
%$b$ are fit parameters. (b)~Time dependent average domain size for five
%different particle concentrations: $C=0.08$ (upper line), $C=0.12$, $C=0.16$,
%$C=0.20$, and $C=0.24$ (bottom line). Every line corresponds to a circle
%in \picref{ggsfCfit_mr1_3v}.}
%\end{figure}

\section{Conclusion and outlook}
\label{secconclusion}
We have demonstrated simulations of anisotropic ellipsoidal particles
stabilizing fluid-fluid interfaces based on a combined multicomponent lattice
Boltzmann and molecular dynamics approach. We provided a description of the
code implementation and its recent improvements enabling our code for
efficient large-scale \RY{simulations} of Pickering/bijel systems on hundreds
of thousands of cores on an IBM Blue Gene/P system while sampling physical
observables such as density fields or colloidal particle configurations with
sufficient temporal resolution. After basic studies of adsorption trajectories
for single ellipsoids we demonstrated that ellipsoidal particles can lead
to a transition between bijels and Pickering emulsions 
depending on contact angle, particle volume concentration or fluid ratio.
Further, we demonstrated that the average lateral domain size of bijels
depends on the particle concentration and can be fitted by a simple $1/C$
relation where the prefactor depends on the particle aspect ratio demonstrating that particles with $m>1$ are more efficient emulsion stabilizers than spheres.

\section*{\RY{Acknowledgments}}
This work was financed by NWO/STW (Vidi grant of J. Harting), by the FOM/Shell
IPP (09iPOG14) and within the DFG priority program ``nano- and microfluidics''
(SPP1264). We thank the J\"ulich Supercomputing Centre for the technical
support and the CPU time which was allocated within a large scale grant of the
Gauss Center for Supercomputing.

%\bibliography{bib/LB}

\end{document}